\documentclass[aps,pre,twocolumn, superscriptaddress, longbibliography]{revtex4-2}
\pdfoutput=1
\usepackage{graphicx}
\usepackage{comment}
\usepackage{grffile} 
\usepackage{dcolumn}
\usepackage{bm,amssymb,amsmath}
\usepackage[pdftex]{hyperref}
\hypersetup{colorlinks=true,citecolor=blue,linkcolor=red,urlcolor=blue}
\usepackage[all]{hypcap}
\usepackage[absolute,overlay]{textpos}
\usepackage{color}
\definecolor{red}{rgb}{0.8, 0.0, 0.0}
\definecolor{blue}{rgb}{0.06, 0.2, 0.65}
\definecolor{green}{rgb}{0,0.6,0}

\usepackage{hyperref}

\begin{document}

\title{Transmutation-accelerated sampling method for multi-component ZrCu(Al) metallic glasses}
\author{Filip Kaśkosz}
\affiliation{NOMATEN Centre of Excellence, National Center for Nuclear Research, ul. A. Soltana 7, 05-400 Swierk/Otwock, Poland.} 
\author{Rene Alvarez-Donado}
\affiliation{Univ. Lyon, INSA-Lyon, CNRS, LaMCoS, UMR 5259, Villeurbanne, 69621, France.}
\affiliation{Laboratoire de Mécanique des Solides, CNRS, Ecole Polytechnique, Route de Saclay, Palaiseau, 91128 France.}
\author{Mikko Alava}
\affiliation{Aalto University, Department of Applied Physics, PO Box 11000, 00076 Aalto, Espoo, Finland.}
\author{Anshul D. S. Parmar}
\affiliation{NOMATEN Centre of Excellence, National Center for Nuclear Research, ul. A. Soltana 7, 05-400 Swierk/Otwock, Poland.} 
\author{Silvia Bonfanti}
\email{Silvia.Bonfanti@ncbj.gov.pl}
\affiliation{NOMATEN Centre of Excellence, National Center for Nuclear Research, ul. A. Soltana 7, 05-400 Swierk/Otwock, Poland.} 

\begin{abstract}
We investigate multi-component metallic glass systems using a hybrid Molecular Dynamics (MD) and Variance-Constrained Semi-Grand Canonical approach. This method enables us to generate samples with properties consistent with experimental observations, at deeply supercooled states that are typically inaccessible with conventional MD simulations.
Using a realistic interatomic potential, we investigate the dynamics, kinetic stability, and rheology of a ZrCu(Al) metallic glass, together with the widely studied ZrCu system, in the low-temperature glassy regime.
Our results demonstrate how the hybrid method enhances relaxation and provides a generic framework for modeling realistic complex metallic glasses.
\end{abstract}

\maketitle
\section{Introduction}

Metallic glasses (MGs) are an important class of materials with an amorphous atomic structure, formed by the rapid quenching of metallic liquids~\cite{greer1995metallic,inoue2000stabilization,wang2004bulk}.  
This disordered structure imparts exceptional mechanical properties, including high strength and a large elastic limit, making MGs highly attractive for technological and industrial applications~\cite{inoue2000stabilization,ashby2006metallic,schroers2010processing}.  
However, large-scale applications remain limited by an incomplete understanding of their glass-forming ability, the structural signatures governing their dynamics, and their long-term thermal and mechanical reliability~\cite{johnson1999bulk}.  
Recent advances in vapor-deposition techniques have enabled the fabrication of glasses equilibrated beyond the experimental glass-transition temperature ($T_g$), i.e., ultrastable states. These states extend the accessible supercooling range for studying and designing glasses~\cite{yu2013ultrastable,luttich2018anti,luo2018ultrastable,sun2020transition,zhao2022ultrastable}.  
Unlike conventional glasses, ultrastable materials exhibit enhanced thermodynamic and kinetic resilience, together with superior mechanical performance~\cite{swallen2007organic,kearns2009high,lyubimov2013model,rodriguez2022ultrastable}. Their stability is said to be comparable to that of liquid-quenched glasses aged for thousands of years~\cite{sun2021film}. Extending the accessible range of supercooled states is therefore of great importance for developing a deeper understanding of glass-forming systems and the existence of putative transitions, which remain major open questions~\cite{debenedetti2001supercooled,dyre2006colloquium}.  

Molecular simulations provide a powerful tool to investigate glasses at the microscopic level~\cite{binder2011glassy,ding2014soft}.  
However, conventional simulations suffer from unrealistically high cooling rates ($\sim [1-100]$ K/ns), producing computer-generated MGs with properties that differ significantly from experiments~\cite{ashwin2013cooling,ryltsev2016cooling,zhang2015cooling,vollmayr1996properties,liu2007cooling}.  
Recently, these timescale limitations have been partially overcome with the development of the Swap Monte Carlo (SMC) method~\cite{grigera2001fast,gutierrez2015static,ninarello2017models,berthier2023modern}, which enables the preparation of well-equilibrated glass samples even below the glass transition temperature $T_g$~\cite{berthier2017configurational,ozawa2018configurational,berthier2019configurational}.   
Pairwise swap moves dramatically accelerate relaxation dynamics, suppressing
crystallization and enabling the exploration of deeply supercooled equilibrium liquids. This strategy is applicable for polydisperse systems, 
but also to carefully designed finite-component Lennard–Jones mixtures, although their efficiency depends strongly on composition~\cite{parmar2020ultrastable}.
In a similar effort to access deeper regions of the energy landscape, other approaches have been proposed. Examples include grand-canonical Monte Carlo with continuous particle size changes~\cite{brito}, augmented potential energy formulations~\cite{kapteijns2019fast}, and thermomechanical annealing protocols~\cite{leishangthem2017yielding,das2022annealing}, all of which extend equilibration or provide access to deeper supercooled states. These developments have significantly advanced the ability to simulate glass formers at low temperatures and motivated the search for further methodologies.  
To account for chemical specificity, hybrid methods combining molecular dynamics (MD), Monte Carlo methods, and realistic potential for metals, e.g. Embedded Atom
Method (EAM)~\cite{daw1984embedded}, have been developed. 
An example from the field of crystalline multi-component alloys, the Variance-Constrained Semi-Grand-Canonical (VC~SGC) ensemble~\cite{sadigh2012scalable}, enables efficient transmutation moves that can accelerate the sampling in the chemical configuration space.
Recently, the VC SGC technique has been applied to binary Zr$_{50}$Cu$_{50}$ metallic glasses, revealing that shear transformation zones are limited to small clusters of particles~\cite{zhang2022shear}, {and highlighting the correlation between mechanical properties and thermodynamic stability~\cite{feng2025correlation}}.
However, this method has so far been restricted to binary MGs, without addressing multicomponent glasses.  

In this work, we propose a generic approach for obtaining realistic ZrCu-based metallic glass samples \textit{in silico}, using an efficient hybrid MD+VC~SGC methodology. Our study focuses on two well-studied metallic glass systems: binary Zr$_{50}$Cu$_{50}$ and ternary Zr$_{46}$Cu$_{46}$Al$_{8}$~\cite{yu2010electronic,cheng2009atomic,ding2014charge,pauly2009phase,luo2018ultrastable}.  
We demonstrate that this approach allows for a systematic study of how variations in elements of specific nature and their concentrations affect the properties of metallic glasses and that it is directly applicable to multi-component systems.  
We obtain samples of metallic systems, examine the dynamics and kinetic stability of their deeply supercooled states, and compare their mechanical properties. Our findings establish a generic methodology for modeling multi-component MGs, applicable to a wide range of atomic species.

\section{Simulation methods}
The atomic interactions were simulated with the EAM potential for ZrCu(Al) glasses~\cite{cheng2009atomic}. 
We performed simulations consisting of $N$=4000 atoms in a cubic box with periodic boundary conditions. For mechanical deformations, we studied samples with $N$=100000 particles. Our simulations were performed with LAMMPS~\cite{lammps}, using a time step $\Delta t$=2~fs. The glass state was obtained by quenching in the isobaric-isothermal ensemble ($NpT$) by integrating the Nos\'e-Hoover equations with damping parameters $\tau_T$=0.2~ps and $\tau_p$=2~ps for the thermostat and barostat. All results were obtained keeping the external pressure $p$=0 bar. We saved configurations during the cooling process for further analysis of the dynamics, thermal stability, and mechanical response.\\
\indent\textit{Hybrid Molecular Dynamics---} 
The hybrid scheme under the VC~SGC ensemble \cite{sadigh2012scalable} is governed by two parameters $\varphi$ and $\kappa$ that are Lagrange multipliers associated with constraints in the first and second moments of the concentration, respectively. The method is based on particle class transmutations and therefore works differently from polydisperse swap, where particle diameters are changed~\cite{brito2018swap}. The hybrid MD+VC~SGC scheme enables the exploration of the configurational degrees of freedom by randomly selecting an atom and attempting to change its type, while also calculating the corresponding energy and concentration changes. Acceptance of these transmutations follows the Metropolis criterion, ensuring the preservation of detailed balance~\cite{sadigh2012excess}. On the other hand, the relaxation processes are taken into account by MD integration steps. 
{Following each transmutation attempt, short MD trajectories are performed to relax local structure, preventing artificial heating and ensuring physically realistic chemical ordering~\cite{sadigh2012scalable,berthier2019efficient}.}

The target ensemble-averaged concentration~$c$ and its fluctuation~$\Delta c$ are controlled by the two parameters mentioned above, $\kappa$ and $\varphi$, according to the relations~\cite{sadigh2012scalable}:
\begin{equation}
\begin{split}
   & \langle c \rangle = (\Delta \mu-\varphi)/2\kappa N,\\
   & \langle \Delta c^2 \rangle \propto  1/\sqrt{\kappa},
\end{split}
\end{equation}
where $\Delta \mu$ is the difference in the chemical potential of two types of particles. In order to avoid an overdrift from the target concentration, the concentration fluctuations should be reduced by increasing the value of $\kappa$. However, as reported in~\cite{sadigh2012scalable}, increasing $\kappa$ results in a reduction in the acceptance rate. Therefore, our strategy was based on determining the difference in chemical potentials that gives the desired composition and choosing a value of $\kappa$ that is a compromise between the acceptance ratio and the fluctuation in the composition. We adopted the value $\kappa = 10^3$ employed by the authors of the method~\cite{sadigh2012scalable} and for the binary ZrCu MGs investigated in~\cite{zhang2022shear}.

\begin{figure}[]
	\centering
	\includegraphics[width=\columnwidth]{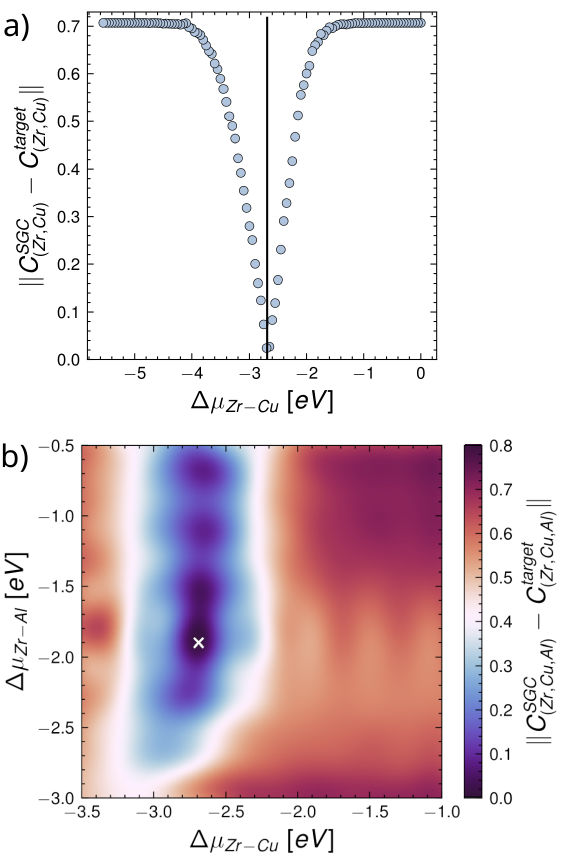}
    \caption{
        The deviation $\lVert C^{SGC}-C^{target} \rVert$ of the simulated composition in SGC from the target {(a)} Zr$_{50}$Cu$_{50}$ and {(b)} Zr$_{46}$Cu$_{46}$Al$_{8}$. The optimal differences in chemical potential $\Delta\mu$ for the target compositions are those that minimize this deviation. For the Zr$_{50}$Cu$_{50}$, the corresponding value is $\Delta\mu_{Zr-Cu}=-2.69$~eV, indicated by vertical black line. For the Zr$_{46}$Cu$_{46}$Al$_{8}$, the optimal values are $\Delta\mu_{Zr-Cu}=-2.69$~eV, and $\Delta\mu_{Zr-Al}=-1.9$~eV, marked by a white cross in the two-dimensional plot. The colorbar represents the deviation, from low (blue) to high (red).}
	\label{figure1}
\end{figure}

First, to determine the chemical potential differences $\Delta \mu$ required to achieve the target compositions for the binary system and the ternary, we employed the Semi-Grand Canonical (SGC) ensemble. Unlike the Variance-Constrained version, which limits concentration fluctuations, the regular SGC ensemble samples system states at fixed chemical potential differences, allowing the composition to vary freely. 

This property enabled us to identify the chemical potential values that yield the target equilibrium compositions. 
At a temperature of $2000$~K, where the systems were in the liquid form, we performed a series of SGC+$NpT$ simulations varying the values of $\Delta\mu$. Each simulation lasted long enough to reach equilibrium, yielding a specific composition $C^{SGC}$, the concentrations of the elements constituting the systems. For the Zr$_{50}$Cu$_{50}$ system, a single chemical potential difference, $\Delta \mu_{Zr-Cu}$, was sufficient to describe the system. 
We varied $\Delta \mu_{Zr-Cu}$ from 0 to -5.5 eV.

\section{Results}

\indent\textit{The optimal chemical potential ---}
Figure~\ref{figure1}(a) shows the Euclidean distance $\lVert C^{SGC}-C^{target}\rVert$ that quantifies the deviation of the simulated composition from the target, as a function of $\Delta \mu_{Zr-Cu}$ for the system Zr$_{50}$Cu$_{50}$. The smallest deviation was observed at $\Delta \mu_{Zr-Cu}=-2.69$~eV, which was subsequently used in the VC~SGC simulations for the next part of the study. 
For the ternary ZrCuAl system, two differences in chemical potentials $\Delta\mu_{Zr-Cu}$ and $\Delta\mu_{Zr-Al}$ are required (the third one can be determined from these two). We explored a grid of values in the range of 0 to -3.5 eV for both, while the intermediate values were interpolated, as shown in Fig.~\ref{figure1}(b). The optimal values $\Delta\mu_{Zr-Cu}=-2.69$~eV, and $\Delta\mu_{Zr-Al}=-1.9$~eV, corresponding to the target composition Zr$_{46}$Cu$_{46}$Al$_{8}$ are marked with a white cross. 
The colormap (third dimension) represents the deviation, from low (blue) to high (red). 
These optimal values of the chemical potential were further utilized for the study.

We performed hybrid MD+VC~SGC simulations for the optimized $\Delta \mu$ values, in which $N$ atomic transmutation attempts were made every 10~MD time steps. We applied an isobaric-isothermal ($NpT$) quenching protocol to 20 samples, starting from a liquid at 1500~K, and cooling to 300~K at a rate of $10^{11}$~K/s. Throughout the process, system configurations were saved for further analysis.
The cooling procedure was systematically interrupted every 50~K by a 0.5~ns thermalization step, during which the potential energy was averaged.

\begin{figure}[thp]
	\centering
	\includegraphics[width=\columnwidth]{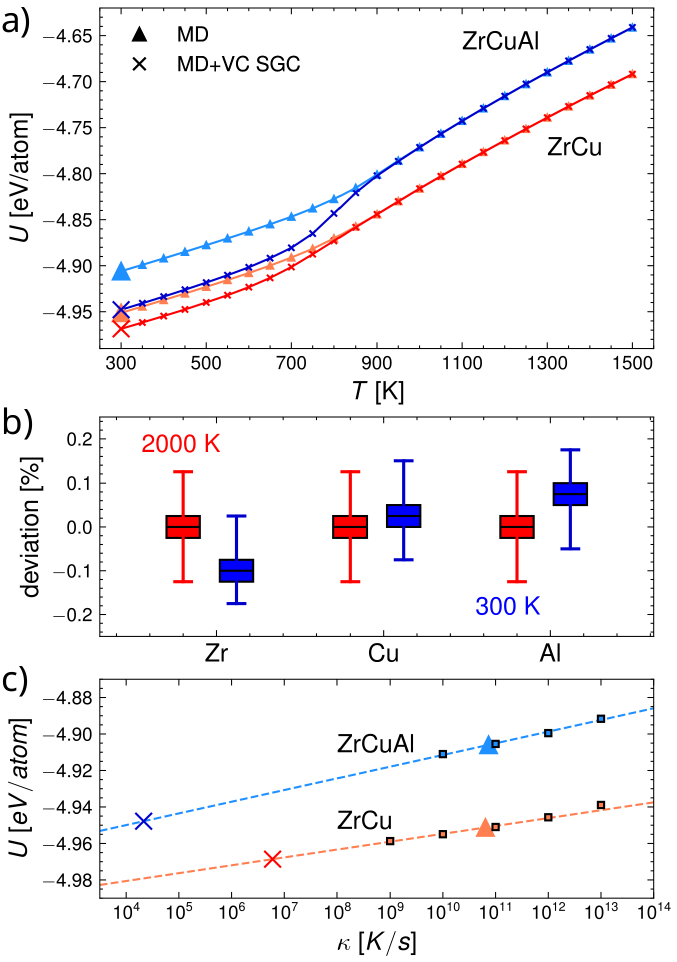}
	\caption{(a) Temperature dependence of the potential energy per atom for the binary Zr$_{50}$Cu$_{50}$ (red lines) and the ternary Zr$_{46}$Cu$_{46}$Al$_{8}$ (blue lines), obtained from cooling simulations using both conventional MD and hybrid MD+VC~SGC methods. The dependencies overlap at high temperatures, where atomic dynamics are fast. As the temperature decreases, the trends diverge, indicating that the hybrid approach enables the system to access lower-energy states more effectively. {(b) Deviations in atomic concentrations from the target composition during the hybrid simulation of ternary ZrCuAl at the highest and lowest temperatures considered in the study (2000 K and 300 K). The composition is preserved as concentration differences for all species are below 0.2\%. (c) Effective cooling rate estimation for hybrid simulations based on a logarithmic dependence of the potential energy on cooling rate, obtained from conventional MD simulations.}}
	\label{figure2}
\end{figure}

For comparison, we performed the same quenching protocol with conventional MD simulation with the same cooling protocol. 
At higher temperatures, where atomic mobility is high, the potential energy dependencies for conventional MD and hybrid MD+VC SGC closely overlap. However, at lower temperatures, the trends begin to diverge, indicating that MD+VC~SGC facilitates enhanced sampling, allowing the system to access lower-energy states (see Fig.~\ref{figure2}(a)). This low-temperature divergence in energy is systematically larger in the ternary system compared to the binary system, suggesting a stronger enhancement of relaxation effects in more compositionally complex systems.

{During hybrid cooling, the system composition remains preserved over the entire temperature range, with atomic concentration deviations below 0.2\%, as shown for ternary ZrCuAl in Fig.~\ref{figure2}(b)
To estimate the effective cooling rate in the hybrid simulations, we performed isobaric cooling using conventional molecular dynamics at rates of $10^{9}$–$10^{13}$ K/s, decreasing the temperature from 1500 K to 300 K. Fig.~\ref{figure2}(c) shows the resulting potential energies at 300 K for the range of cooling with conventional MD simulation, suggesting the potential energy exhibits a logarithmic dependence over the cooling rate~\cite{zhang2022shear}. For the hybrid simulations, we estimate the effective cooling rates with the logarithmic extrapolation of the obtained potential energies, yielding approximately $10^{7}$ K/s for ZrCu and $10^{4}$ K/s for ZrCuAl, reflecting that the hybrid method provides an efficient approach to access deeper supercooled states.}

\begin{figure}[]
	\centering
	\includegraphics[width=\columnwidth]{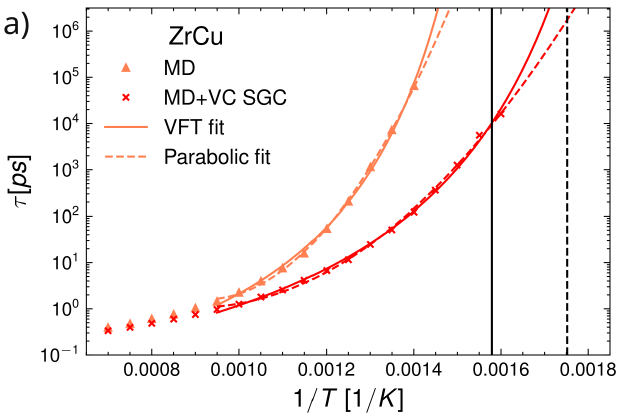}
	\includegraphics[width=\columnwidth]{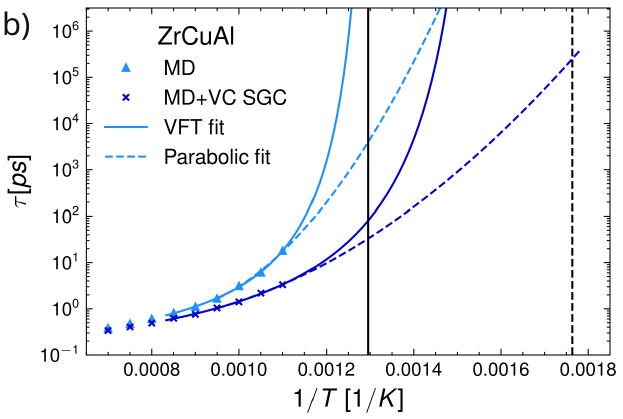}
        \caption{
        Relaxation times obtained from conventional MD and hybrid MD+VC-SGC simulations as a function of inverse temperature for (a) the binary and (b) the ternary systems. The solid and dashed lines represent fits to the Vogel–Fulcher–Tammann (VFT) and parabolic forms, respectively. The glass transition temperature $T_g$ was estimated by extrapolating the fitted MD relaxation times to 100~s. From the VFT fit, $T_g = 633$~K for ZrCu and $T_g = 772$~K for ZrCuAl, as indicated by the vertical solid lines. From the parabolic fit, $T_g \approx 570$~K for both systems, indicated by the vertical dashed lines. The hybrid approach systemically provides relatively faster sampling. }
        \label{figure3}
\end{figure}

\indent\textit{Dynamics ---}
We now compare the dynamical behavior of the systems, in the equilibrium, using both simulation approaches. First, the configurations obtained during the hybrid cooling were subjected to further MD+VC~SGC simulations at constant pressure and temperature, allowing the systems to reach equilibrium. 
The simulations were sufficiently long for the intermediate incoherent scattering function (IISF), evaluated at the wave vector corresponding to the maximum of the static structure factor~\cite{kaskosz2022}. Production runs were then carried out to determine the relaxation times ($\tau$), defined as the time at which the IISF decays to $1/e$. The resulting equilibrated configurations were subsequently used as initial samples for conventional MD simulations, from which relaxation times were similarly extracted. {We emphasize that the relaxation times obtained from the hybrid MD+VC SGC simulations quantify structural-decorrelation and serve for comparison with the standard MD for the computational efficiency, therefore should not be interpreted as intrinsic physical timescales.}

{At sufficiently low temperatures, we observed crystallization in both systems using both conventional and hybrid MD simulations. In the binary system, crystallization was observed at $T = 690$ K, for approximately 50\% of the samples (during hybrid MD+VC SGC), relaxation times were therefore determined exclusively from trajectories of the system that remained liquid. For the ternary system, with the higher sampling efficiency, at $T= 870$ K, all analyzed samples obtained from the hybrid approach exhibited crystalline ordering. The system could not relax to a metastable-supercooled state, so the relaxation time could not be determined. The crystallization behavior is consistent with previous reports for this EAM potential~\cite{cheng2009atomic,ryltsev2018,indicators,one_more_cheng_icosa} and with other studies that agree with experimental observations~\cite{icosahedral_AIMD, icosahedral_RMC,ORTIZ2025121402}.
We also present a detailed structural analysis based on the local bond-orientational order parameter $q_6$~\cite{steinhardt1983}, which shows that crystallization occurs in both conventional and hybrid MD simulations (see Supplementary Material (SM), Sec.~I~\cite{supp,steinhardt1983,rosenfeld1998density}). In the liquid regime, however, both methods sample thermodynamically and structurally equivalent states (see SM, Sec.~II~\cite{supp,soni_kamal}).}

{A comparison of the relaxation times obtained from both approaches is shown in Fig.~\ref{figure3}(a,b), the panels correspond to the binary and ternary systems, respectively. 
For conventional MD, over the temperature range spanning approximately one to five decades of structural relaxation time in the supercooled regime, both systems exhibit a non-Arrhenius temperature dependence. We do not observe a distinct fragile-to-strong crossover within this limited supercooling window, in contrast to the experimental findings of Zhou \textit{et al.}~\cite{zhou2015structural}, which were reported over a much broader temperature range. Accordingly, we describe the extrapolated dynamics and the glass transition using both Vogel–Fulcher–Tammann (VFT)~\cite{Floudas2011} and parabolic fits~\cite{elmatad}, which together capture the behavior and sampling efficiency over the accessible simulation regime.}

{The glass transition temperature $T_g$ was estimated by extrapolating the MD relaxation times to a value of 100~s. Using the parabolic fit, we obtain $T_g^{\mathrm{para}} \approx 570$~K for both systems, while the VFT extrapolation yields higher values, $T_g^{\mathrm{VFT}} = 633$~K and $772$~K for ZrCu and ZrCuAl, respectively. Figure~\ref{figure3}(a,b) shows the corresponding parabolic and VFT fits, with the estimated glass transition temperatures indicated by vertical lines. At $T_g^{\mathrm{para}}$, the hybrid MD+VC-SGC method samples configurations approximately eight orders of magnitude faster than conventional MD for both systems, whereas using $T_g^{\mathrm{VFT}}$, the corresponding speedups are roughly 8 and 10 orders of magnitude for ZrCu and ZrCuAl (in the absence of crystallization), respectively. We note that both VFT and parabolic extrapolations are sensitive to the chosen supercooling range. Nevertheless, the hybrid approach consistently enables efficient sampling of equilibrium states, while the quantitative speedup depends on the accessible temperature window.}

\begin{figure}[]
	\centering
	\includegraphics[width=\columnwidth]{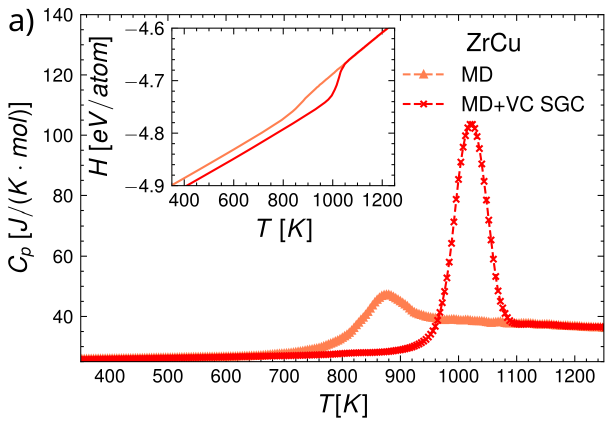}
	\includegraphics[width=\columnwidth]{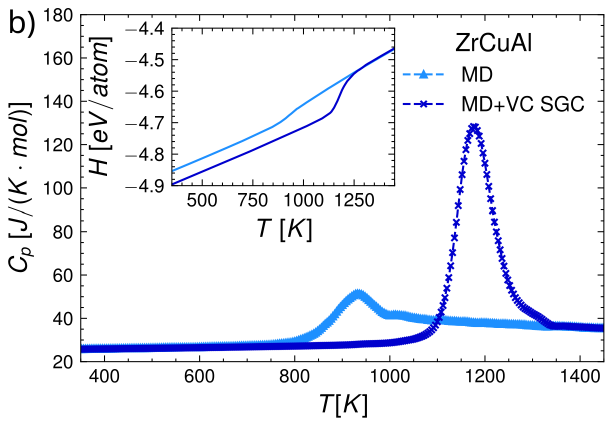}	\caption{Heat capacity as a function of temperature determined during the regular MD heating {($10^{11}$ $K/s$)} of {{(a)}} Zr$_{50}$Cu$_{50}$ and {{(b)}} Zr$_{46}$Cu$_{46}$Al$_{8}$ samples resulting from MD and MD+VC SGC cooling procedures. The insets show the temperature dependence of the enthalpy per atom obtained from heating, whose derivative was used to determine the heat capacity.}  
	\label{figure4}
\end{figure}

\indent\textit{Kinetic stability.---}
Inspired by experiments on metallic glasses produced via vapor deposition, a common method to assess their kinetic stability is through calorimetric measurements during controlled heating. In Fig.~\ref{figure4}, we compare the thermal behavior of glasses prepared using conventional MD and those produced with the hybrid MD+VC~SGC approach. The glass samples were heated using standard molecular dynamics at a rate of $10^{11}$ K/s, and the temperature dependence of enthalpy $H(T)$ was analyzed, see the insets of Fig.~\ref{figure4}. The results were averaged over 20 independent samples. To smooth the enthalpy curves, a moving average was applied. The specific heat was determined using the relation $C_p = (dH/dT)_p $, and the resulting curves are presented in the main panel. {At this point, it is worth noting that the heating rates accessible in atomistic simulations ($\sim 10^{9}$–$10^{12}$ K/s) are many orders of magnitude higher than those used in experiments (typically $1$–$40$ K/min, i.e., $\sim 10^{-2}$–$10^{0}$ K/s)~\cite{smith2017separating}. Consequently, the enthalpy recovery occurs over a narrower temperature interval, leading to sharper and higher $C_p$ peaks compared to experiments.} However, it is evident that glasses produced using the hybrid MD+VC~SGC method exhibit the onset of the melting process at higher temperatures, indicating significantly enhanced kinetic stability compared to those prepared via conventional MD.

\begin{figure}[]
	\includegraphics[width=\columnwidth]{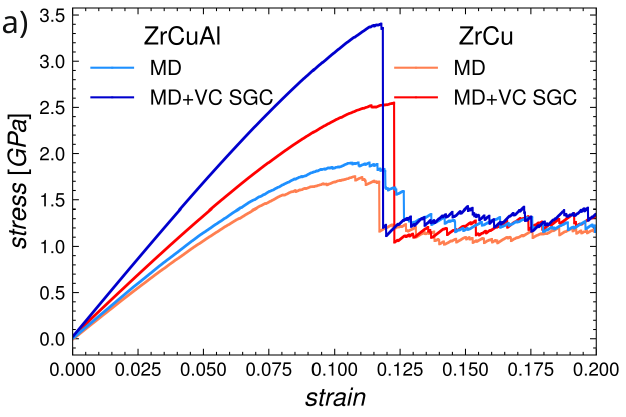}
	\includegraphics[width=\columnwidth]{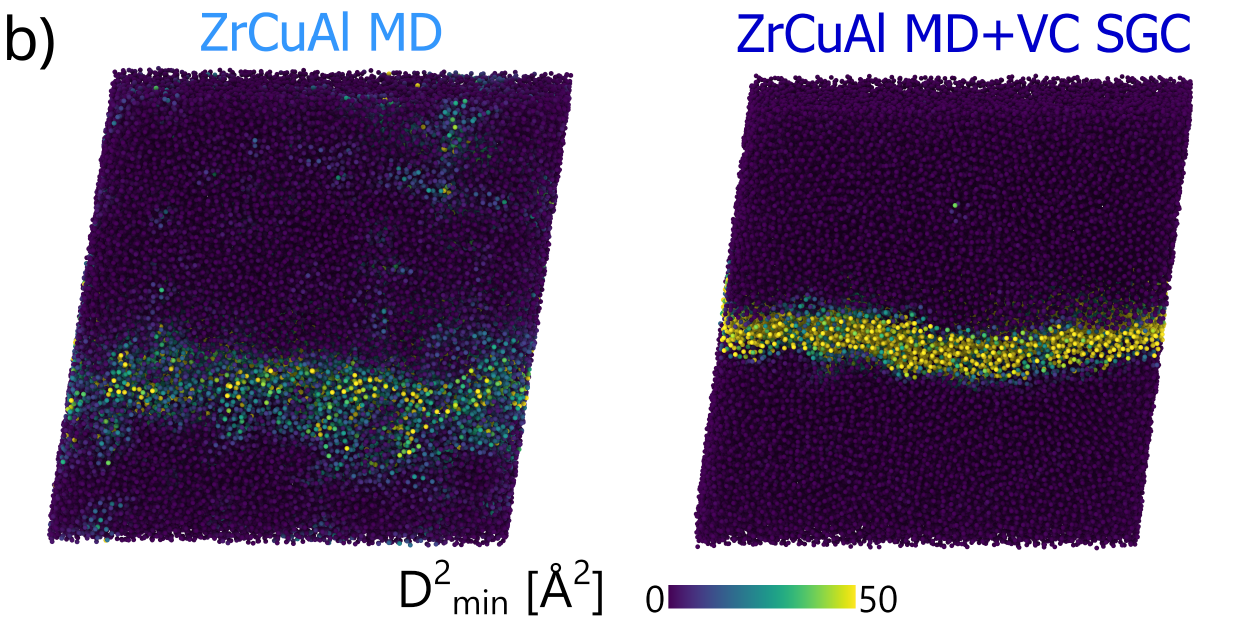}
    \caption{(a)~Stress-strain curves for samples obtained from MD and MD+VC~SGC cooling procedures. Samples prepared with MD+VC~SGC scheme exhibit drastic stress drop. 
    (b)~Example of shear bands for ternary ZrCuAl MG. The shear band is more diffuse for samples prepared with conventional MD simulations (left panel), while it is well-defined for samples generated via MD+VC~SGC (right panel). The colorbar indicates the non-affine squared displacements $D^2_{min}$ from the unstrained configuration, {as an indicator to identify the atoms that have participated in plastic events}.}
	\label{figure5}
\end{figure}
\indent\textit{Rheology.---} 
To assess the mechanical response, we performed athermal quasistatic shear (AQS) simulations~\cite{maloney2006amorphous} on configurations obtained by quenching systems of $N$=100000 atoms to $T$=300~K. The simulation box was sheared incrementally in the $x-$direction using a strain step $\delta\gamma=10^{-4}$. After each increment, energy minimization was carried out using the Fast Inertial Relaxation Engine (FIRE)~\cite{bitzek2006structural} scheme until mechanical equilibrium was reached, defined as a maximum residual force below 10$^{-8}$eV/\AA.
Figure~\ref{figure5}(a) shows the shear stress–strain curves for samples obtained from conventional MD and hybrid MD+VC~SGC cooling procedures. Figure~\ref{figure5}(b) displays the
non-affine squared displacements $D^2_{min}$~\cite{falk1998}, relative to the initial unsheared configurations.
The shear modulus was extracted by performing linear regression on the stress–strain data in the elastic regime (strain range: 0–2\%). For ZrCu, the shear modulus is 22.5~GPa and 27.2~GPa, while for ZrCuAl, it is 23.8~GPa and 33.9~GPa, as determined for samples obtained via conventional MD and hybrid MD+VC~SGC approaches, respectively.
These values represent averages over five independent samples, with standard deviations less than 1\%.
For conventional MD simulations, the stress exhibits an initial increase, followed by tiny plastic events, an overshoot with larger plastic events, and a subsequent approach to the steady state. Near the yield, the strain map is spatially heterogeneous 
resulting in a broad and diffuse shear band~\cite{ozawa2018random} (see Fig.~\ref{figure5}(b), left panel).   
{For samples generated via MD+VC-SGC, the stress–strain curves exhibit a pronounced stress drop prior to plastic flow, corresponding to the formation of a system-spanning shear band (see Fig.~\ref{figure5}(b), right panel), consistent with observations in model glasses~\cite{ozawa2018random,parmar2020ultrastable}. This suggests that the proposed methodology can access states exhibiting both ductile and brittle responses, which is particularly relevant for capturing the mechanical response of realistic glassy systems.}

\section{Summary \& Conclusions} In order to resolve the physics of glasses \textit{in silico} - a big challenge due to the complexity of glasses and their history-dependent formation - novel ideas are needed.
In this work, we show how the hybrid Molecular Dynamics and Variance Constrained Semi-Grand Canonical (MD+VC~SGC) simulation method, as a generic approach, is able to tackle this for multi-component MGs. 
{We present results that clearly show an approach to sample deeper-supercooled states, in an efficient manner.}
Finally, this fact is seen to make a difference in kinetic stability and mechanical properties. 

In conclusion, our results provide a substantial contribution to the field of glass physics, particularly with regard to multi-component metallic glasses (MGs)~\cite{zhang2022recent,bonfanti2023quasi,wadowski2024efficient}. Using advanced simulations to obtain low-$T$ glassy states, we provide an essential methodology for further studies, shedding light on the underlying physical mechanisms that govern the complex behaviors and properties of these amorphous materials. Our work prompts further theoretical investigations aimed at probing the microscopic mechanisms of complex MGs at a fundamental level, such as their relaxation dynamics~\cite{guiselin2022microscopic}. 
{The methodologies employed in this work not only open promising directions toward accessing states closer to experimental conditions, but also suggest new avenues for optimizing the structure of metallic glasses via controlled cooling and processing strategies for a wide range of applications.}

\section*{Code Availability}
The data presented in the figures, as well as the simulation scripts implementing the hybrid procedure, can be found at \url{https://zenodo.org/records/19813564}.

\section*{Acknowledgements}
We gratefully thank Misaki Ozawa for fruitful discussions. FK, ADSP, and SB are supported by the European Union Horizon 2020 research and innovation program under grant agreement no. 857470 and from the European Regional Development Fund via the Foundation for Polish Science International Research Agenda PLUS program grant No. MAB PLUS/2018/8. SB and FK acknowledge support from the National Science Center in Poland through the SONATA BIS grant DEC-2023/50/E/ST3/00569 (PI: SB). SB and ADSP acknowledge support from the Foundation for Polish Science in Poland through the FIRST TEAM FENG.02.02-IP.05-0177/23 project (PI: SB). This work was carried out within the ``Projektowanie Ulepszonych Szkieł Metalicznych" project (FENG.02.02-IP.05-0177/23) under the 2.2 First Team programme of the Foundation for Polish Science co-financed by the European Union from the European Funds for Smart Economy 2021–2027 (FENG).

\appendix

\bibliography{biblio}
\end{document}